\documentclass[11pt,spanish]{article}
\usepackage[T1]{fontenc}
\usepackage[latin1]{inputenc}
\usepackage{geometry}
\usepackage{graphicx}
\IfFileExists{url.sty}{\usepackage{url}}
                      {\newcommand{\url}{\texttt}}

\makeatletter

\providecommand{\tabularnewline}{\\}

\usepackage{babel}
\addto\extrasspanish{\bbl@deactivate{~}}
\makeatother
\begin{document}

\title{El Informe NERA analizado}

\author{Ricardo Galli\\
Universitat de les Illes Balears%
\footnote{Agradezco a los miembros de la lista de Bulma (\protect\url{http://bulmalug.net}),
especialmente a Xosé Otero, Paco Ros, Francico J. Martinez y Giles
A. Radford por sus correcciones y sugerencias de último momento.%
}}

\maketitle
\begin{abstract}
This is a  review of the article
{}``Government Preferences for Promoting Open-Source
Software: A Solution in Search of A Problem\char`\"{} \cite{evans1,evans2}
by David Evans and Bernard J. Reddy.  This report was paid for by Microsoft
and put together at its request. Now Microsoft is using it as part of their
lobbying campaign in Europe against governments' promotion of Open Source Software.
As expected, this article is strongly biased and most of the conclusions
are based upon false hypotheses and evidence.

-----

Este artículo es un análisis crítico del \emph{dossier} de David Evans
y Bernard J. Reddy: {}``Government Preferences for Promoting Open-Source
Software: A Solution in Search of A Problem\char`\"{} \cite{evans1,evans2}.
El informe ha sido realizado por encargo de Microsoft con el objetivo
de usarlo en sus campañas de \emph{lobby} contra el fomento del código
abierto, especialmente el software libre y la licencia GPL. Dicho
informe está siendo entregado a legisladores y miembros de gobiernos,
especialmente europeos, con el objetivo de detener o minimizar la
promulgación de leyes que discriminen positivamente al código abierto.
Como era de esperar, el informe es muy parcial y está lleno de falsas
hipótesis y evidencias.
\end{abstract}

\section{Introducción}

Microsoft está llevando a cabo una inmensa tarea global de \emph{lobbying}
para convencer a los gobiernos y legisladores de que no elaboren leyes
que favorezcan al código abierto (\emph{open source}) y al software
libre, en particular a los liberados bajo licencia GPL. Está tan alarmado
Microsoft con el crecimiento y apoyo social al software libre que
presionó furiosamente%
\footnote{Parece haber ganado, porque al día de hoy la reunión está suspendida
indefinidamente.%
} a través de la ya conocida BSA (\emph{Business Software Alliance})
para anular una reunión de la Organización Mundial de la Propiedad
Intelectual (WIPO) que debía realizarse en Suiza con el objetivo de
debatir acerca de las implicaciones del software libre \cite{key-10}.

Una de las acciones emprendidas por Microsoft es enviar una carpeta
con una serie de informes a los políticos europeos que están vinculados
con acciones legislativas a favor de una discriminación positiva del
código abierto. 

\begin{figure}
\begin{center}\includegraphics[%
  scale=0.3]{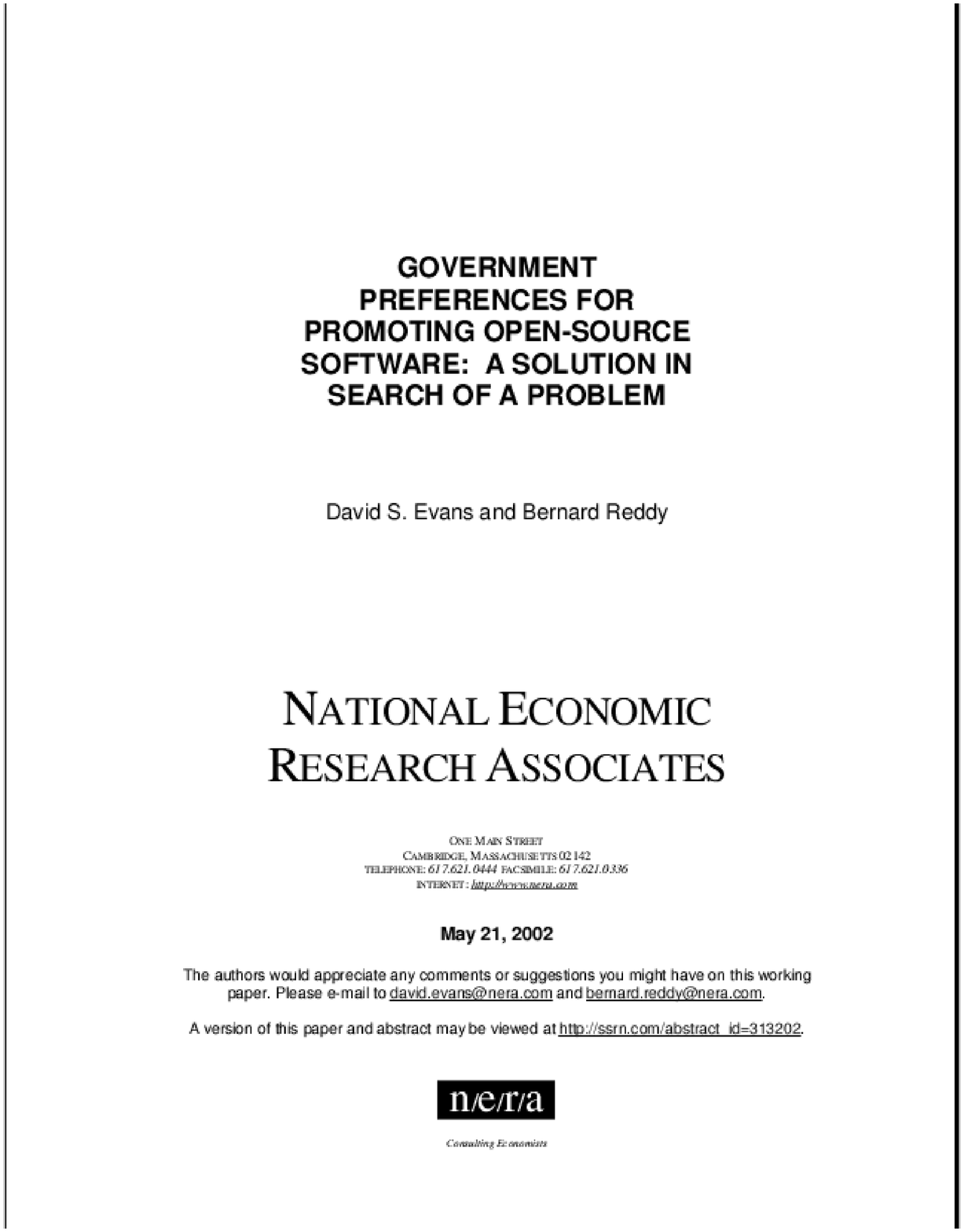}\end{center}

\caption{\label{cap:Informe-Evans-Reddy}Informe Evans-Reddy}
\end{figure}

Uno de los informes más conocidos, o al menos que tiene más impacto,
es el informe titulado \emph{Government Preferences for Promoting
Open-Source Software: A Solution in Search of a Problem}%
\footnote{{}``Preferencia de los gobiernos para promover el software de código
abierto: una solución en búsquedas de un problema''.%
}. La primera versión, y la más divulgada, de este documento es de
mayo de 2002%
\footnote{Curiosamente en las mismas fechas sale publicado otro artículo en
la misma línea titulado \emph{Public Subsidies for Open Source? Some
Economic Policy Issues of the Software Market}, de Klaus Schmidt y
Monika Schnitzer. Los autores agradecen a Evans y Reddy por su ayuda
y a la financiación de NERA.%
} \cite{evans1} (Figura \ref{cap:Informe-Evans-Reddy}). Una segunda
versión ---con cambios mínimos, especialmente para adecuarla mejor
a un formato artículo científico-técnico--- fue publicada en el volumen
9 (invierno de 2003) de \emph{Michigan Telecommunications and Technology
Law Review} \cite{evans2}.

\begin{figure}
\begin{center}\includegraphics[%
  width=0.60\columnwidth]{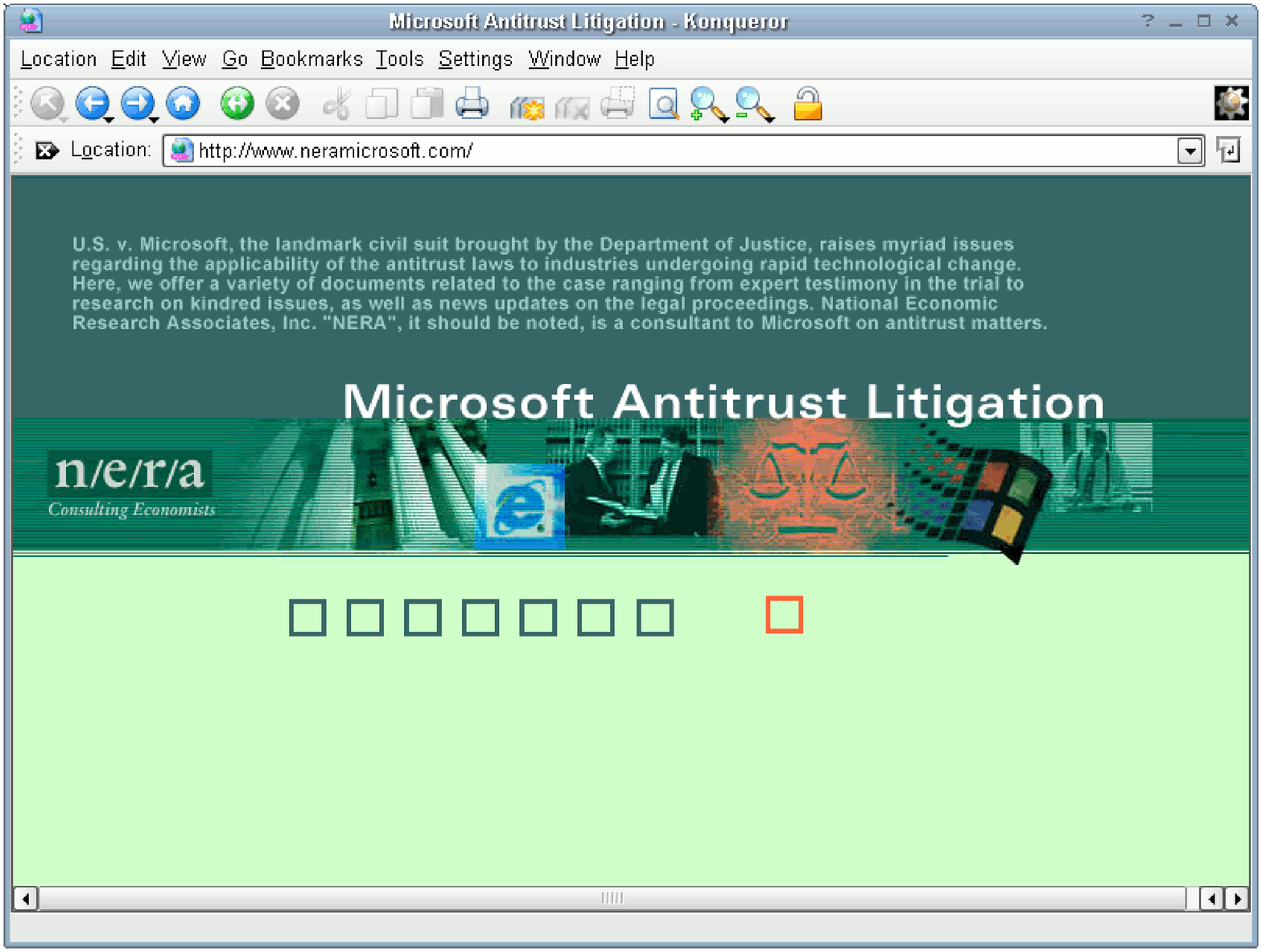}\end{center}

\caption{\label{cap:NeraMicrosoft.com}NeraMicrosoft.com}
\end{figure}

Como se puede observar en la Figura \ref{cap:Informe-Evans-Reddy},
el informe está elaborado por la empresa NERA (\emph{National Economic
Research Associates}) y escrito por David Evans ---vicepresidente
de NERA--- y Bernard Reddy. Aunque el informe tiene la apariencia
y estilo de un estudio independiente, en realidad es un informe pagado
por Microsoft, tal como lo aclaran en la primera nota al pie. Como
ellos mismos explican en \url{www.neramicrosoft.com} (Figura \ref{cap:NeraMicrosoft.com}),
NERA es una empresa que trabaja casi exclusivamente para Microsoft
en áreas económicas y de monopolio. Durante todo el juicio anti-monopolio
del Estado (USA) contra Microsoft, los dos autores han trabajado elaborando
informes y declarando a favor de Microsoft.

Es tan estrecha la relación entre ambas empresas, o la dependencia
de una respecto a la otra, que NERA no ha tenido dudas en ligar el
nombre de Microsoft y su imagen corporativa a la de NERA, como se
puede observar en el sitio web expresamente desarrollado (Figura \ref{cap:NeraMicrosoft.com}).

Es decir, el informe NERA aparenta ser un \emph{dossier} independiente
pero está \emph{escrito a pedido y pagado por Microsoft} con objetivos
muy concretos: convencer a los legisladores de que cualquier norma
que favorezca al {}``código abierto'' es perjudicial para la economía%
\footnote{Aunque no especifica si es malo para la economía del país del legislador,
para la economía de Estados Unidos o para la economía de Microsoft
y NERA.%
}.

\section{Razonamiento y conclusiones del informe NERA}

El documento está redactado con un estilo innecesariamente confuso
pero aparentemente bien fundamentado, sobre todo por el uso excesivo
de términos técnicos económicos%
\footnote{Lo que los anglo-sajones llaman \emph{mumbo jumbo}.%
}, y la repetición de suposiciones e interpretaciones subjetivas para
que parezcan hechos comprobados%
\footnote{La licencia GPL es mala, el {}``código abierto'' no es innovador,
la industria es muy innovadora...%
}. Desde el punto de vista de la estructura, el documento está bastante
desbalanceado y mal organizado ---hasta en la numeración de las secciones---
que hace que al lector le resulte difícil ubicarse en el contexto.
Aunque pretende ser un informe analítico, en realidad combina hechos
con estadísticas parciales y opiniones para obtener hipótesis que
luego son usadas como verdades irrefutables%
\footnote{Parece que este tipo de razonamiento ya fue descrito por Ramon Llull
como {}``razonamiento circular'' en su \emph{Ars Magna}. Sus estudios
para evitar este tipo de razonamientos le hizo concebir lo que algunos
científicos consideran el antepasado primigenio de la ciencia informática.%
}.

Las conclusiones formales del informe pueden resumirse en:

\begin{itemize}
\item No hay problemas en el mercado que justifiquen la intervención del
gobierno.
\item La industria del software es particular, pero es igual a todas las
industrias basadas en propiedad intelectual.
\item El software propietario ha sido muy innovador y beneficioso para los
usuarios.
\item El código abierto ha tenido algunos éxitos, pero el software GPL ha
tenido muy pocos.
\item Los gobiernos no saben como hacerlo, por lo tanto no deben intervenir
en el mercado del software.
\item No creemos en las {}``externalidades''%
\footnote{\emph{Efecto red}. Explica el informe que gracias al efecto red los
usuarios se benefician del monopolio de Microsoft, por ejemplo, todos
los usuarios pueden intercambiar ficheros Word. Pero si el gobierno
interviene para crear aplicaciones distintas ese beneficio se diluiría
y perjudicaría a los usuarios. Obviamente intentan confundir {}``aplicación''
con {}``formato de intercambio de datos''. Si hubiese diversidad
de aplicaciones, además de disminuir riesgos de propagación de virus
y de dar libertad a los usuarios para elegir la {}``aplicación'',
Microsoft finalmente empezaría a hacer los filtros que hace años dejó
de desarrollar. O empezaría a documentar perfectamente sus formatos
propietarios o confusos y variables.%
}, pero si los gobiernos las consideran es mejor que inviertan más
en I+D de software, pero que no financien el desarrollo de software
GPL.
\end{itemize}
¿Como llegan a todas estas conclusiones? 

El informe tiene 79 páginas, analizarlo para explicar las contradicciones
en cada párrafo necesitaría otro documento de igual o mayor longitud,
así que nos centraremos en las líneas que consideramos más importantes
del informe.

Aunque no está exactamente organizado de esta forma, las ideas centrales
son:

\begin{enumerate}
\item La industria informática no está concentrada.
\item La calidad del software se incrementó, pero los precios bajaron.
\item Por \textbf{1} y \textbf{2} no existen problemas en el mercado.
\item Durante los 90 se incrementaron el número de patentes de software
registradas.
\item Debido a \textbf{4}, hay innovación en el mercado del software.
\item La licencia GPL es mala para los negocios y
\item el {}``código abierto'' no es innovador, aún menos el software desarrollado
con licencia GPL.
\item Los gobiernos no deben tomar decisiones políticas, sólo técnicas y
económicas.
\item Por \textbf{3}, \textbf{5}, \textbf{6}, \textbf{7} y \textbf{8} los
gobiernos no deben financiar a proyectos de software GPL.
\end{enumerate}
Varios de los hechos e hipótesis que presentan son ---como mínimo---
discutibles. Seguramente los lectores iniciados en temas de software
libre, patentes y monopolios informáticos ya se están haciendo las
siguientes preguntas:

\begin{itemize}
\item Con la evidencia del monopolio de Microsoft%
\footnote{93\% del Windows, entre 90\% y 95\% del Microsoft Office.%
}, ¿cómo se puede justificar que la industria informática no está concentrada?
\item Con el número creciente de virus que se propagan rápidamente%
\footnote{MSBlast y SoBig en menos de dos semanas. Además de los problemas de
seguridad de los sistemas afectados se debe al {}``efecto red'':
la mayoría de los usuarios usan la misma aplicación.%
} por errores de diseño y programación de programas básicos como el
navegador de Internet o el lector de correo electrónico más utilizado,
¿cómo pueden decir que la calidad del software se ha incrementado
notablemente en los últimos años?
\item Con un tema tan complejo y controvertido como es el de las patentes
de software y la relación entre ellas y la innovación. Con los estudios
y análisis de científicos reconocidos que presentan evidencias contrarias%
\footnote{\url{http://ffii.org}%
} ¿cómo pueden extrapolar directamente el aumento de patentes de software
con mayor innovación? ¿No podría ser que antes no existían las patentes
de software ---o eran más complicadas de obtenerlas%
\footnote{Los criterios se han relajado a principios de los 90.%
}--- y por eso en los últimos años se han incrementado?
\item Si la licencia GPL es relativamente joven pero ha permitido el desarrollo
de una cantidad inmensa de software ---disponibles de forma gratuita---,
además con nuevas empresas que basan su negocio en software liberado
con esta licencia ¿cómo pueden asegurar que la GPL es mala para los
negocios?, ¿cuántas pequeñas y medianas empresas ---hasta grandes
empresas como Google, O'Reilly y Amazon--- tienen toda su infraestructura
funcionando con software libre y Linux?
\item Si Internet, el correo electrónico, los estándares web, los sistemas
operativos avanzados como Unix, editores de texto, compiladores, lenguages
de programación, etc. se han desarrollado en entornos abiertos y de
{}``compartir el código fuente'', respetando estándares y siguiendo
los principios científicos fundamentales como la revisión de los pares,
la construcción acumulativa sobre los avances anteriores, ¿cómo afirman
y justifican que el {}``código abierto'' no es innovador?%
\footnote{Hay estudios que aseguran que las tecnologías actuales fueron desarrolladas
en gran parte durante los 80.%
}
\item Si la tarea y responsabilidad fundamental de los cargos políticos
electos es precisamente tomar decisiones políticas para asegurar el
bienestar general ¿cómo se puede afirmar que los gobiernos no saben
y no deben tomar decisiones políticas sobre industrias y tecnologías
que afectan a todas las actividades de un país desarrollado?. ¿Cómo
una empresa norteamericana ---condenada en su propio país por actividades
monopolistas predatorias%
\footnote{Y con otro juicio y posible multa pendiente en la Unión Europea.%
} en el área informática--- pretende que los gobiernos europeos no
legislen sobre el mismo tema para evitar las desigualdades y actividades
predatorias? ¿Cómo pueden sugerir que los gobernantes deben tomar
decisiones como si fuesen empresas? Es decir, sólo maximizar ganancias
y minimizar gastos%
\footnote{Página 67, en la crítica a Lessig.%
}.
\end{itemize}

\section{Análisis detallado}

En las siguientes seis secciones se elaboran análisis más detallados
de lo que consideramos el núcleo importante del informe NERA. 

En la sección \ref{sub:Hay-m=E1s-patentes,} se analiza si realmente
el software propietario ha sido fuente de tantas innovaciones en los
últimos años, y sobre todo si esa innovación está demostrada por el
hecho de que se hayan incrementado el número de patentes de software.

En la sección \ref{sub:No-hay-problemas} se analiza si tal como dice
el informe NERA, realmente existe poca concentración en la industria
del software y el porqué de la contradicción de esa supuesta baja
concentración con el monopolio de Microsoft en los sistemas y aplicaciones
de escritorio.

El informe asegura que en los últimos años la calidad del software
ha aumentado y los precios han caído, sobre todo gracias al {}``monopolio
natural'' de Microsoft. Lo analizamos en la sección \ref{sub:La-calidad-del}.

En la sección \ref{sub:GPL-es-mala} se analiza la falsedad ---repetida
a lo largo de todo el informe NERA--- de que la licencia GPL es intrínsecamente
mala para los negocios.

Así como el informe asigna la innovación como característica casi
exclusiva del software propietario, también argumenta la falta de
innovación existente en el software libre. Estos últimos son analizados
en la sección \ref{sub:El-c=F3digo-abierto}.

Finalmente la sección \ref{sub:El-gobierno-no} analiza las exigencias
para que los gobernantes no tomen decisiones políticas, ni intervengan
en los mercados, ni financien proyectos de desarrollo de software
GPL. Y si lo hacen, que sea BSD.

\subsection{\label{sub:Hay-m=E1s-patentes,}Hay más patentes, por lo tanto más
innovación}

El estudio de Microsoft (página 7, 20), establece una relación directa
entre el número de patentes de software y la innovación, investigación
y desarrollo: {}``como en la década de los 90 hubo un incremento
de registros de patentes de software, entonces hay más innovación
y la industria es floreciente''. Dado que existen estudios e informes
que indican lo contrario, como mínimo debería considerarse dicha afirmación
como controvertida. 

El informe compara el número de patentes registradas en Estados Unidos
durante la década de los 80 con las registradas en los 90. La diferencia
es considerable, y de ello infieren que la innovación se ha incrementado.
Pero no mencionan que los \emph{criterios de patentabilidad se han
relajado considerablemente a principios de los 90}. El más significativo
de los cambios fue la modificación de las pruebas para determinar
la obviedad de una invención, los nuevos criterios toman más en cuenta
factores secundarios ---como el éxito comercial--- para calificar
de {}``no obvia'' a una invención \cite{hunt}. 

Es como eliminar las pruebas prácticas de conducción para obtener
la licencia de conducir y años después ---ante el incremento más que
probable del número de licencias otorgadas--- inferir que existe una
mejora de la educación vial de la población. 

La relación entre patentes e innovación es un tema tan complejo y
controvertido, que hasta Alan Greenspan ---presidente de la Reserva
Federal de Estados Unidos--- tiene dudas de cuál es el sistema adecuado
para proteger las invenciones más basadas en las ideas que en objetos
físicos \cite{key-12}. 

Carl Shapiro, especialista reconocido en temas de patentes, también
se cuestiona y pregunta si el sistema legal de patentes norteamericano
está correctamente diseñado para promover la investigación mas que
desalentarla \cite{shapiro}. Shapiro también menciona que importantes
investigaciones están demostrando la tendencia de las compañías a
registrar nuevas patentes y elaborar estrategias muy agresivas%
\footnote{Como \emph{patent mining} o \emph{submarine patents.}%
}. El informe de Shapiro acaba afirmando que el sistema de patentes
norteamericano está llevando a una situación potencialmente peligrosa,
especialmente en los campos de biotecnología, semiconductores, software
informático y comercio electrónico.

Es especialmente relevante un estudio muy reciente de Bessen y Hunt
\cite{patents}. Según dicho estudio, si las patentes promoviesen
la investigación se podría demostrar que son complementarias a la
I+D. Sin embargo el estudio demuestra que en la década de los 90 el
\emph{registro de patentes ha pasado a ser un sustituto de la I+D}.
Claramente las \emph{patentes de software no son un incentivo (o actividad
complementaria) de la investigación y desarrollo}.

Según Bessen y Hunt, las patentes de software en USA significan el
15\% de todas las patentes, la mayoría adquiridas por grandes empresas
y de manufacturación. 

\begin{itemize}
\item El sector de manufacturación posee el 69\% de las patentes de software,
pero sólo emplea al 10\% de los programadores. 
\item Por otro lado, todas las empresas del sector de {}``publicación de
software y servicios'' (que incluye a IBM) solo poseen el 16\% de
las patentes de software, pero emplea al 42\% de los programadores
y analistas. 
\end{itemize}
Ello coincide con la opinión de la mayoría de ejecutivos de empresas
de software que afirman que las patentes tienen poco valor para la
actividad de sus empresas%
\footnote{Hay que destacar que William Neukom ---de Microsoft--- fue uno de
los que más ha defendido la influencia positiva de las patentes de
software para la investigación e innovación: \url{http://www.jamesshuggins.com/h/tek1/software_patent_microsoft.htm}%
} \cite{hearing}.

Las patentes de software pueden haber sido complementarias de la I+D
en la década de los 80, cuando los estándares para obtener patentes
eran relativamente altos, pero eso ha cambiado totalmente en la década
de los 90, cuando las patentes se han convertido en sustitutas de
la investigación y desarrollo. Según los cálculos del estudio, \emph{la
actividad de I+D hubiese sido hasta un 10\% ó 15\% más alta sin la
sustitución de la investigación y desarrollo por las estrategias de
patentes}. 

El \emph{mayor uso de patentes de software está asociado con intensidades
más bajas de I+D}. 

Las patentes de software han sido registradas principalmente por empresas
que se dedican a construir {}``matorrales de patentes''%
\footnote{\emph{Patent thickets}%
}, situación que preocupa especialmente a Shapiro \cite{shapiro} por
las dificultades de negociación de patentes a la hora de introducir
un nuevo producto al mercado.

Finalmente, aunque los autores del Informe NERA utilizan el recurso
de contabilizar números de patentes para convencer que la innovación
florece en la industria del software, más adelante (página 54) ellos
mismos ponen en entredicho todo el sistema de patentes de software
cuando resumen los problemas expuestos por tres premios Nobel de Economía%
\footnote{George Akerlof, Joseph Stiglitz, Michael Spence.%
}:

\begin{quotation}
El sistema de patentes es un método imperfecto del gobierno para remediar
esos fallos {[}falta de diseminación del conocimiento{]} ---el gobierno
asegura a los inventores un monopolio temporal sobre sus invenciones
a cambio de una \emph{exposición completa}%
\footnote{Cabe recordar que el software propietario no obedece ---porque no
está obligado--- a la exposición o revelado del secreto ---código
fuente--- para obtener una patente.%
} {[}de las invenciones{]}.
\end{quotation}
Puede afirmarse entonces que no hay evidencias de que el sistema de
patentes de Estados Unidos sea el adecuado. No hay evidencias de que
ese mismo sistema {}``relajado'' de patentes de software sea complementario
a las actividades de I+D en la industria del software. Por lo tanto
tampoco se puede deducir que la industria del software esté en una
época de innovación floreciente basados solamente en un análisis cuantitativo
de patentes. Probablemente está ocurriendo exactamente lo contrario.

\subsection{\label{sub:No-hay-problemas}No hay problemas en el mercado informático}

Desde el inicio los autores del informe dejan clara su postura respecto
de la intervención de los gobiernos en los mercados. Ya en el segundo
párrafo de la página 2 exponen: 

\begin{quotation}
En los últimos 20 años, la mayoría de los gobiernos han elegido incrementar
su confianza en las fuerzas del mercado para gobernar la producción
y distribución de bienes y servicios.%
\footnote{Lo expresado por los primeros autores está en concordancia con la
evolución de la {}``Escuela de Chicago'', que hasta la década de
los 80 defendió la intervención de la FCC para evitar los monopolios
como el de AT\&T, pero ahora siguen los argumentos de {}``integración
vertical como forma de optimizar las 'externalidades'''. David Evans
explica en \cite{wp} que la división de AT\&T en las {}``Baby Bells''
fue beneficiosa porque no innovaba ---se olvidó de la invención del
transistor de los laboratorios Bell, los premios Nobel, o del desarrollo
del mismo UNIX---. En coherencia con la {}``post Escuela de Chicago''
ahora defiende lo contrario: el monopolio de Microsoft es beneficioso.%
}
\end{quotation}
Seguramente se refieren, entre otras cosas, a la liberalización de
los años 80 promovida por la \emph{Federal Communications Commission}
(FCC) de USA. O al proyecto de Ley (\emph{Billtext}) HR776 de 1992
del Congreso Norteamericano para incrementar la competencia y mejorar
la {}``eficiencia de la producción y distribución de energía''.
Hechos recientes, como los cortes generalizados de luz en todo el
nordeste norteamericano \cite{lights}, escándalos como Enron ---la
principal productora de energía de USA---, o el gran fracaso de ofrecer
telefonía móvil celular compatible y con cobertura global, tal como
se hizo en Europa \cite{gsm}%
\footnote{Notar además que uno de los autores de este artículo, David Salant,
es de la misma consultora NERA, y ha realizado el estudio por encargo
de la misma. Aunque el estudio intenta justificar la necesidad de
{}``no intervención'' del gobierno en la telefonía móvil.%
} --- que luego de casi 10 años todavía lleva una ventaja considerable
a USA--- plantean serias dudas de las capacidades auto-reguladoras
de las {}``fuerzas del mercado'' y de los beneficios de las decisiones
liberalizadoras de infraestructuras tan básicas.

Respecto a la situación de la industria informática, en la página
12 se afirma que no existe una concentración importante en la industria
del software y que los índices de concentración de otras industrias
es superior a la informática. No solamente se contradice con la sentencia
del juicio en USA%
\footnote{Y la percepción y sentido común de los usuarios.%
} (\emph{mantenimiento ilegal de monopolio, actos y prácticas excluyentes}),
sino que en la página 73 del informe se afirma que en el año 2000
\emph{Microsoft ha distribuido el 93\% de los sistemas operativos
clientes mono-usuario}%
\footnote{Los que conocemos generalmente como ordenadores personales, hogareños
o de escritorio. Aunque comete un error en denominarlos {}``mono-usuarios''.
Técnicamente el Windows NT, 2000, XP y 2003 son ordenadores multi-usuario.%
}.

La falacia consiste en tomar y comparar datos estadísticos muy genéricos
y de dos fuentes distintas. En primer lugar usan datos de índices
de concentración del Censo de Estados Unidos de 1997 (nota al pie
44). Usan lo que llaman los cuatro dígitos SIC, en particular el 5112,
que corresponde a la {}``Publicación y Reproducción de Software''.
La oficina del Censo incluye en esta rama industrial a todas las empresas
que se dedican a la producción, distribución, documentación, asistencia
de instalación y servicios. Es decir, están incluidas todas aquellas
empresas, grandes, pequeñas y hasta unipersonales, cuyas actividades
estén relacionadas a cualquier rama de la informática, desde la programación,
servicios de consultoría y mantenimiento de aplicaciones y copia de
datos hasta la publicación de manuales o documentación de programas.
En pocas palabras, si los \emph{top-mantas} fueses legales, seguramente
estarían contabilizados en este capítulo. 

Primero toma los datos del Censo de Estados Unidos ---las del código
5112---, pero la oficina del censo norteamericana no proporciona índices
de concentración de la industria informática, el menos las correspondientes
al censo de 1997. Para poder comparar el nivel de concentración de
la industria del software con otras industrias toman los datos otro
informe de características muy disímiles, el informe de IDC (\url{www.idc.com})
que está elaborado con técnicas y objetivos distintos a los del censo. 

El documento de IDC referenciado (\#25569) no pudo ser encontrado
en el web de IDC, sin embargo se encontró uno muy similar y que posiblemente
sea la versión actualizada del informe \cite{idc}%
\footnote{No hemos podido leer dicho informe debido a su elevado coste ---4.500
dólares---. %
}. Las notas de Evans-Reddy y el resumen de \cite{idc} disponible
en el web indican que la muestra para obtener los índices HHI%
\footnote{El índice HHI \emph{---Herfindahl-Hirschman Index---} se calcula de
la siguiente manera. Si hay tres empresas, cada una de ellas con una
penetración porcentual de mercado A, B y C respectivamente, donde
$A=40,B=35,C=25$, el índice HHI será igual a:

\[
HHI=40^{2}+35^{2}+25^{2}=1600+1225+625=3.450\]

Si se consideran 10 empresas, cada una de ellas con una participación
de mercado del 10\%, el HHI será:

\[
HHI=10^{2}+10^{2}+...+10^{2}=10*10^{2}=1.000\]

Si ahora se quiere calcular el HHI para una industria de 100 compañías,
cada una con un 1\% del mercado:

\[
HHI=100*1^{2}=100\]

Se observa que el índice será más alto mientras mayor sea la concentración
de mercado en menos empresas (el máximo es 10.000).%
} del informe de IDC incluyen sólo a las cien principales empresas
productores de software de características y mercados muy diferentes.

Según el informe de Evans y Reddy, el informe de IDC indica que el
HHI de las firmas de {}``paquetes de software'' es igual a 244.
Luego hace una comparación de los índices de IDC con los datos de
la Oficina del Censo de Estados Unidos%
\footnote{Cabe recordar que la Oficina del Censo sólo publica los índices de
concentración para empresas de manufacturación, no existen datos para
la industria del software.%
} y que comprende a otras industrias mucho más específicas, como los
automóviles (2.506) o los cereales... ¡de desayuno! (2.446). 

\emph{Como resultado de esa comparación de fuentes y metodologías
distintas, deducen que la industria informática tiene unos índices
de concentración muy bajos} por lo que no hay deficiencias en el mercado
(¿y por lo tanto Microsoft no es un monopolio?). 

Pero en la página 73 indican que Microsoft distribuyó en el año 2000
el 93\% de sistemas operativos de escritorio, una penetración de mercado
inexistente en ningún sector importante de la industria. Haciendo
cálculos rápidos ($93^{2}$) observamos que el \emph{HHI en este sector
es como mínimo 8.469}.

La otra aplicación importante de Microsoft ---que le genera una parte
importante de sus ingresos--- es la \emph{suite} ofimática Microsoft
Office. Este conjunto de aplicaciones, que incluye el procesador de
texto Word, la hoja de cálculo Excel y el programa para presentaciones
PowerPoint, están presentes en los ordenadores Windows, ya sean personales
o de la administración, de tal forma que los formatos propietarios
de los ficheros de esos programas se usan para intercambio de datos,
hasta en los ficheros que la administración pone a disposición de
los usuarios. Estudios independientes \cite{office1,key-7} revelan
que la \emph{suite} Office de Microsoft tiene entre un 94 y 95 por
ciento del mercado, los que significa un índice \emph{HHI de por lo
menos 8.836}.

Según el mismo informe (nota al pie número 46), si el índice es superior
a 2.000 suele ser una preocupación para el Departamento de Justicia
y la Comisión Federal de Comercio de Estados Unidos. 

Por los elevados índices que acabamos de mostrar, queda claramente
reflejado que \emph{las áreas de {}``sistemas operativos de escritorio''
y de {}``}suites \emph{ofimáticas'' están altamente concentradas,
y ambos mercados monopolizados por la misma empresa}. 

Los datos comentados demuestran que la situación del mercado de la
informática no puede considerarse en ningún caso como {}``normal'',
y seguramente existen problemas de monopolios y actividades anti-competitivas.

\subsection{\label{sub:La-calidad-del}La calidad del software aumentó y los
precios bajaron}

Los autores del informe NERA recurren repetidamente (páginas 16, 17,
18, ...) al argumento de que los precios ---en valores nominales y
{}``ajustados a la calidad''--- han decrecido continuamente. Por
un lado explican: 

\begin{quotation}
... la calidad del software ha crecido enormemente en los últimos
años. 
\end{quotation}
Esa afirmación, desde el punto de vista de la ingeniería del software
es como mínimo discutible. El concepto {}``calidad'' está interpretado
en el informe como {}``el número de características o funcionalidades''
del software, no como una medida cualitativa de los componentes ni
como medida cuantitativa de la disminución de errores o disfunciones
de los programas. 

Por otro lado no toman en cuenta que la elaboración de {}``precios
ajustados a la calidad'' es muy compleja con el software y aún más
con los sistemas operativos y \emph{suites} ofimáticas. No se pueden
separar claramente en componentes y características para comparar
sus precios individuales, como se hace en la {}``metodología o índices
hedónicos'' \cite{arthur,hugo}.

Los argumentos de precios ajustados y las referencias citadas corresponden
a argumentos presentados por Microsoft en el juicio \cite{juicio,little}.
Sin embargo se cuidan de no citar las audiencias, pruebas y documentación
presentadas por la Federación de Consumidores de América (CFA) \cite{cfa}.
Por ejemplo en la prueba número 1 (Tabla \ref{cap:Precios-de-sistemas})
muestran los precios comparativos de los sistemas operativos pre-instalados
y el precio de venta en las tiendas. Esta tabla pone en evidencia
que el \emph{monopolio de Microsoft no ha favorecido la reducción
de precios de sistemas operativos}.

\begin{table}
\begin{center}\begin{tabular}{|c|c|c|}
\hline 
\multicolumn{3}{|c|}{Precios de sistemas operativos}\tabularnewline
\hline
\hline 
&
OEM&
Precio en tiendas\tabularnewline
\hline 
1981&
\$ 40&
ND\tabularnewline
\hline 
1990&
\$ 19&
\$ 138\tabularnewline
\hline 
1998&
\$ 50&
\$ 179\tabularnewline
\hline
\end{tabular}\end{center}

\caption{\label{cap:Precios-de-sistemas}Precios de sistemas operativos (Fuente:
CFA)}
\end{table}

\begin{figure}
\begin{center}\includegraphics[%
  scale=0.5]{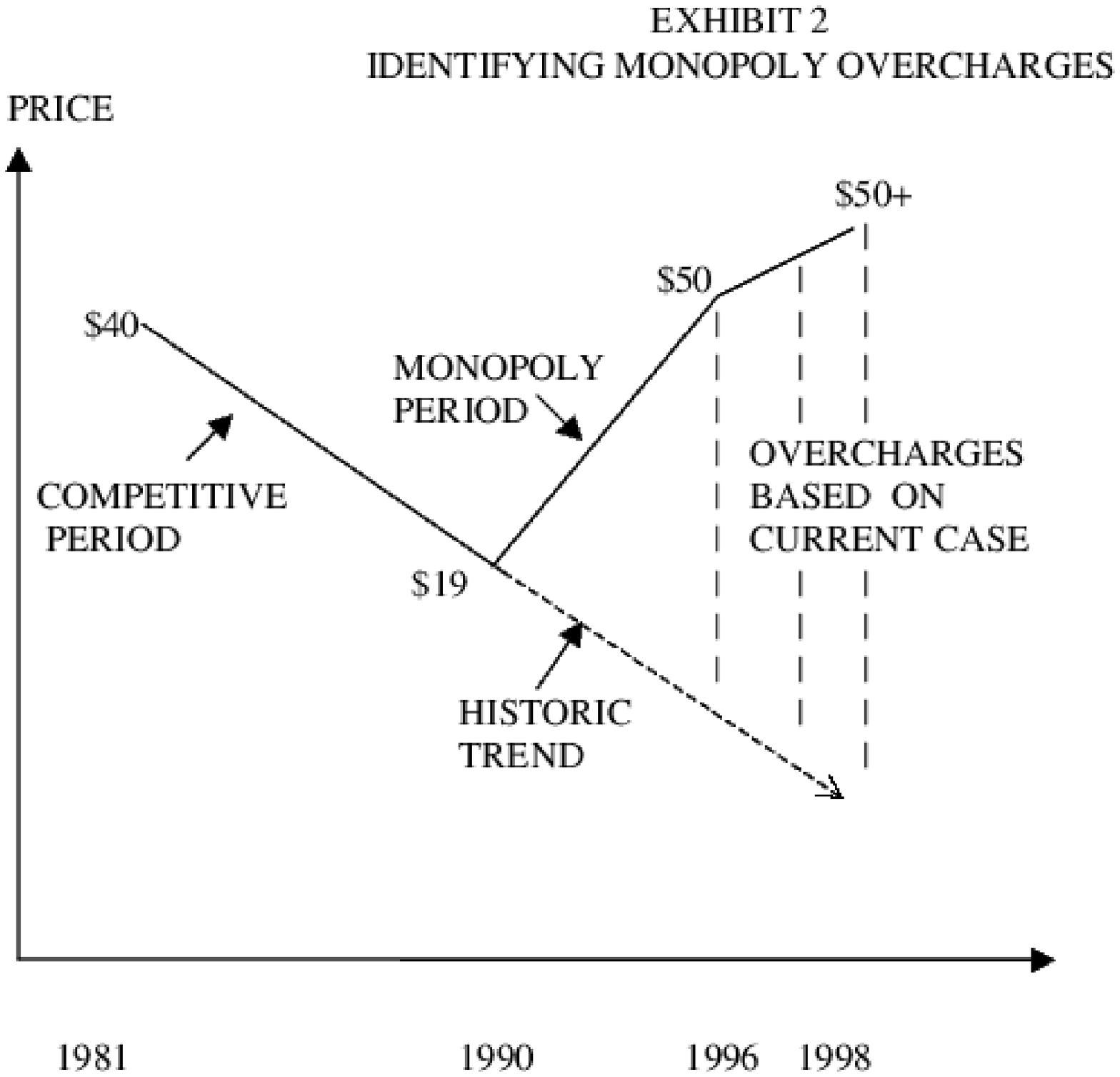}\end{center}

\caption{\label{cap:Sobreprecios-monopol=EDsticos}Sobreprecios debidos al
monopolio (Fuente: CFA)}
\end{figure}

La CFA analizó la evolución de los precios antes y después del monopolio
de Microsoft y demuestra que sí existieron sobreprecios favorecidos
por el monopolio de Microsoft (Figura \ref{cap:Sobreprecios-monopol=EDsticos}).

En otro informe de Evans y Reddy \cite{little} se hace un estudio
para justificar el precio de Windows como {}``competitivo''. Entre
los parámetros se indica que el hardware cuesta unos 2.000 dólares,
mientras que el Windows menos de 100. Esas proporciones, al menos
actualmente, no se mantienen. Por ejemplo el Windows XP Professional
(OEM) cuesta en una tienda 169 euro, el {}``MS Office XP Professional'',
399 euro. Por otro lado, un PC de muy altas prestaciones, con pantalla
TFT incluida se ofrece a menos de 1000 euro (todos con IVA incluido)
casi en cualquier tienda informática.

\subsubsection{Los juegos de consolas también cuentan}

El informe también menciona precios de productos que no tienen relación
con los sistemas operativos y aplicaciones de escritorio, por ejemplo
el de los videojuegos de consolas.

En la página 8 mencionan precios de aplicaciones y utilidades informáticas,
pero incluyen en la comparativa a un juego de la consola Sony Play
Station 2 (\emph{Final Fantasy X}). Dicha inclusión indica un desconocimiento
profundo de las diferencias de metodologías, diseño, desarrollo, grupos
de trabajo y tecnologías entre desarrollar una {}``aplicación informática''
y juegos comerciales. Algunas de las características diferenciadoras
del desarrollo de juegos son:

\begin{itemize}
\item Ciclo de vida: los juegos tienen un ciclo de vida muy corto, hay que
comercializarlo en muy poco tiempo. Mientras que el desarrollo puede
durar varios años, la comercialización se realiza en un período muy
corto de tiempo, no superior a 2 años.
\item Presupuesto: aunque el ciclo de vida es relativamente muy corto, los
presupuestos de desarrollo son enormes, similares a la producción
de una película de cine, e involucran grupos interdisciplinarios y
muy especializados.
\item Arte, argumento, gráficos: el desarrollo de aplicaciones informáticas
corrientes consiste básicamente en la especificación de los algoritmos
mediante un lenguaje de alto nivel. En los juegos sólo una pequeña
parte del trabajo (y del presupuesto) se dedica a las actividades
de programación, el resto es escribir la historia, diseñar los árboles
de navegación, desarrollar la parte gráfica, con un \emph{componente
artístico considerable} (y costoso en tiempo y dinero). Gran parte
de ese desarrollo artístico debe ser realizado desde cero para posibles
secuelas o nuevas versiones de los juegos.
\item Sustitución del código: mientras que en las aplicaciones informáticas
el código es acumulativo, los juegos suelen ser en gran parte sustitutos.
Ya sea porque toda la parte artística, gráfica, argumental es nueva,
o porque las tecnologías de juegos avanzan muy rápidamente.
\end{itemize}
Aún con todas las diferencias descritas, en la página 40 el informe
vuelve a insistir en comparar la industria del software y los sistemas
operativos con la de los juegos ---y de paso insistir en las supuestas
deficiencias del software con licencia GPL---:

\begin{quotation}
No tenemos conocimiento de ningún juego de calidad comercial desarrollado
bajo la licencia GPL.
\end{quotation}
A pesar de las diferencias y que quizás no cumplan sus {}``requerimientos
de calidad'', podríamos citar Xpilot, GNU Go, GNU Chess, el paquete
GCompris, Tuxracer, Armagetron, Nethack y Freeciv como ejemplos de
que también existen juegos con licencia GPL ---algunos de ellos muy
populares en la comunidad---.

\subsection{\label{sub:GPL-es-mala}La licencia GPL es mala para los negocios}

En la página 10 se menciona ---por primera vez--- que la licencia
GPL puede tener un coexistencia difícil con el software propietario.
Aunque en algunos casos puede ser verdad ---por ejemplo si Microsoft
desea modificar e integrar módulos GPL \emph{completos} en sus propios
programas propietarios--- pero no es la regla para el 95\% de las
empresas que desarrollan software pero que no lo comercializan de
la misma forma que lo hace Microsoft%
\footnote{Se estima que un 95\% de los programas desarrollados son para uso
internos (\emph{in-house}) de las empresas.%
}.

El Linux tiene licencia GPL, pero no existe ningún problema legal
para desarrollar, compilar, probar, ejecutar y comercializar aplicaciones
comerciales propietarias que se ejecuten sobre Linux. De hecho, Microsoft
podría adaptar y comercializar, si desearan, todas sus aplicaciones
para Linux gracias a dos razones fundamentales:

\begin{enumerate}
\item La interacción entre las aplicaciones y el núcleo del sistema operativo
se realiza a través de una interfaz perfectamente definida ({}``llamadas
de sistemas'') y las aplicaciones que la usan no son consideradas
parte ni trabajo derivado del núcleo del sistema operativo.
\item La licencia LGPL \cite{lgpl}. Esta licencia permite que se compilen
e incluyan funciones ({}``edición de enlaces'') de librerías bajo
licencia LGPL en programas propietarios y/o con licencias incompatibles
con la GPL sin que haya conflictos. La LGPL fue diseñada y definida
específicamente para evitar estos problemas. Dicha licencia suele
ser usada en todas las librerías fundamentales del sistema GNU/Linux,
por ejemplo en las libc (biblioteca de funciones elementales e imprescindibles
de un sistema Unix) o en las librerías gráficas GNU/Gnome.
\end{enumerate}
También hay empresas que publican y comercializan sus programas con
licencias duales, GPL ---si se usan para programas compatibles con
dicha licencia--- u otra licencia distinta ---generalmente comercial
y similares a las licencias comerciales de productos propietarios---
si. Tal es el caso de las librerías gráficas QT de Trolltech o de
bases de datos MySQL.

En la página 39 se afirma:

\begin{quotation}
El {}``código abierto'' ha sido incapaz de desarrollar software
amigable para el mercado de masas.
\end{quotation}
Ya hay estudios que indican claramente que la facilidad de uso de
los escritorios Gnome y KDE son igual de simples (o igual de complejos)
de usar que las últimas versiones del sistema Windows \cite{usa-kde,usa2,usa3}. 

En la misma página 39 afirman que la única \emph{suite} informática
de código abierto exitosa es {}``probablemente'' OpenOffice, y a
continuación que {}``es la base de un producto propietario, el StarOffice''.
No sólo se contradicen en que el código abierto es {}``perjudicial
para los negocios'' dando un ejemplo contrario sino que además no
mencionan otra \emph{suite} informática que está avanzando rápidamente,
KOffice.

También se han olvidado de mencionar otros productos exitosos de {}``mercados
de masa'', tales como el Evolution, Mozilla, Konqueror, Kmail, proyecto
Kolab%
\footnote{Resultado del \emph{Kroupware Project}, un proyecto comercial realizado
para el \emph{Bundesamt für Sicherheit in der Informationstechnik}
(Agencia Federal Alemana de Seguridad de Tecnologías de la Información).
\url{http://kolab.kde.org}%
}, etc. Es notable además que los componentes de visualización Html
(Khtml) y ejecución de Javascript (Kjs) del KDE, ambos con licencia
GPL, han sido seleccionados para el desarrollo del navegador web de
Apple, Safari\cite{safari}. Es otro ejemplo de una empresa importante
de desarrollo de software que se beneficia del software libre, con
licencia GPL, y sin que existan problemas legales. 

El Linux y software GPL en general han servido de plataforma para
otros proyectos innovadores muy importantes, como el primer grabador
doméstico de vídeo digital, TiVo (\url{www.tivo.com}). Ante los requerimientos
tan complejos de interacción con vídeo digital y tiempo real ---grabación
y reproducción simultánea--- han seleccionado al Linux y herramientas
GNU para su plataforma.

Las intenciones de NERA de transmitir información parcial quedan nuevamente
desveladas en la página 32 y la nota al pie 101. En ambas explica
que Corel había dejado de dar soporte a Linux y que la última versión
de su producto estrella ---Wordperfect--- en Windows es más reciente
que la de Linux. Pero no informan que en Octubre de 2001 Microsoft
hace una {}``alianza estratégica'' con Corel \cite{key-8}, con
una inversión estimada en 150 millones de dólares%
\footnote{Y bajo presión por amenazas de demandas de Microsoft a Corel por violación
de patentes.%
}. Es la propia empresa Microsoft que ha forzado a Corel a dejar de
desarrollar sus productos para Linux. Seguramente una estrategia más
para mantener su monopolio en el área de las \emph{suites} ofimáticas.

\subsection{\label{sub:El-c=F3digo-abierto}El {}``código abierto'' no es innovador,
aún menos la GPL}

En diversas partes del informe se menciona la falta de innovación
en el {}``código abierto'', especialmente del software libre con
licencia GPL. La primera de este tipo de afirmaciones ---en la página
20--- marca una clara diferencia entre la época anterior al ordenador
personal (desintegrada, programas caros o menos versátiles) versus
la era del PC, donde el software propietario ha florecido y con él
han bajado los precios y aumentado la flexibilidad.

Pero no mencionan que el PC no es producto de la {}``innovación''
de Microsoft, ni siquiera de IBM, sino que fue concebido por un grupo
de entusiastas%
\footnote{\emph{Hackers}%
} ---miembros de lo que en ese tiempo era el {}``código abierto''---
que, aprovechando la potencia del recién creado microprocesador, querían
hacer llegar los ordenadores a cada hogar \cite{homebrew1,homebrew2}.
El grupo más importante de esos entusiastas se hacía llamar \emph{Homebrew
Computer Club} y estaba formado entre otros por Steve Wozniac, uno
de los fundadores de Apple.

Durante ese periodo {}``pre PC'' se han definido las características
principales de los sistemas operativos, las mismas que ha usado Microsoft
más de 10 años después para implementar el Windows 95 y Windows NT.
A principios de los 90 toda la tecnología de Internet ya estaba definida
y desarrollada ---tal como lo reconoce el mismo informe NERA--- con
software libre: BIND, Sendmail, FTP (usado por Microsoft en sus productos),
Telnet (también usado por Microsoft). 

También el desarrollo de la {}``Interfaz Gráfica de Usuario'' es
previa a la época de los PCs. Sus orígenes son el sistema y lenguaje
de programación Smalltalk, desarrollado en Xerox Palo Alto. La primera
empresa que comercializó exitosamente ordenador personales con interfaz
gráfica fue Apple, fundada por uno de los miembros del \emph{Homebrew
Computer Club}. 

Cuando Microsoft lanza su primer sistema gráfico como producto comercial
exitoso (Windows 3.11), el sistema de Apple ya llevaba varios años
en el mercado, también existían implementaciones de ese paradigma
de interacción con el usuario en todos los sistemas Unix, principalmente
basados en el estándar X-Window.

En otras palabras, la intervención de Microsoft ---a pesar de su poderío
y dominio del mercado--- no ha definido las características más innovadoras
de la industria, ni de los sistemas operativos, ni de las tecnologías
de Internet, ni de la interfaz hombre-máquina, y ni siquiera de las
\emph{suites} ofimáticas (cuyos orígenes comerciales se sitúan en
el Wordstar, dBaseII, Lotus 1-2-3).

En la página 38 se afirma que el software más conocido con licencia
GPL es Linux pero que es sólo usado mayoritariamente en servidores,
pero no en {}``clientes''%
\footnote{Por sistemas de escritorio.%
}, a pesar que hay estudios que indican que Linux ya tiene una porción
del mercado de ordenadores {}``clientes'' superior a Apple%
\footnote{Las estimaciones varían entre el 2 y 5 por ciento del mercado.%
}. El sistema Linux y las herramientas GNU/GPL se ejecutan eficientemente
en una gran diversidad de plataformas ---por lo menos 11 distintas,
desde grandes servidores y ordenadores de escritorio a sistemas empotrados
y móviles, como TiVo, Zaurus, servidores inalámbricos de Linksys/Cisco---
a diferencia de los productos de Microsoft, que sólo funcionan en
la arquitectura Intel%
\footnote{O \emph{Wintel.}%
}. 

A pesar de las razones recién explicadas ---compatibilidad y flexibilidad---
dice el informe en la página 48 que el Linux no es innovador, sino
{}``imitador''. Pero no se han percatado de un fenómeno inesperado
generado a partir del desarrollo de Linux: \emph{una revolución en
la ingeniería del software}, objeto de estudios de informáticos y
economistas \cite{bazaar,penguin,quality,Ganesh}. 

Hace sólo diez años muy poca gente ---incluso los expertos en sistemas
operativos--- creía que se podría desarrollar algo tan complejo como
un sistema operativo con gente desconocida y sólo comunicándose entre
ellos a través de Internet%
\footnote{La lista de correo \emph{Linux Kernel Mailing List} o LKML. \url{http://lkml.org}%
}, con una coordinación y dirección relativamente baja%
\footnote{Comparado a los estándares tradiciones de desarrollo de software.%
}, sin un diseño previo y sin que haya retribución monetaria directa
a los programadores.

En la página 46 del informe se afirma:

\begin{quotation}
Samba es sólo un intento de clonar las funcionalidades de Windows. 
\end{quotation}
¡Por supuesto! El Samba%
\footnote{http://samba.org%
} es un software para permitir la integración de sistemas Unix con
los métodos y protocolos propietarios de Microsoft. De nada serviría
que se desarrolle un sistema de integración que no implementase las
funcionalidades de una de las plataformas.

A continuación (página 46) dicen:

\begin{quotation}
MySQL es sólo una implementación de una base de datos con características
estándares (SQL). 
\end{quotation}
Otra vez ¡por supuesto!, esta es una de las características fundamentales,
y una de las ventajas y necesidades del código abierto: la compatibilidad
con estándares es necesaria y beneficiosa. La empresa que desarrolla
MySQL nunca ha querido competir en características no estándares con
bases de datos comerciales como Oracle o DB2 de IBM, sólo ser una
base de datos muy eficiente, de bajo coste y mantenimiento. Objetivos
claramente conseguidos, al ser una de las bases de datos más usadas
en sitios web dinámicos.

En la página 47 hay una frase que resume en pocas palabras las contradicciones
del informe:

\begin{quotation}
Hasta que Internet se comercializó, había pocas razones para que las
empresas intentaran escribir software propietario para ella.
\end{quotation}
Internet se considera una de las revoluciones más importantes desde
la imprenta, y como lo manifiestan en el informe NERA, las empresas
de desarrollo de software propietario comercial \emph{no han participado
en el desarrollo de la misma hasta que estuvo preparada para generar
beneficios monetarios a las empresas}. 

Entonces ¿quién ha innovado?.

Una parte considerable del desarrollo que hizo posible Internet ---desde
los protocolos, pasando por el servidores de nombres (BIND), correo
electrónico (Sendmail), hasta servidores web (Cern, httpd, Apache)
y navegadores web (Mosaic, Amaya)--- fue desarrollado en un entorno
de {}``código abierto'' con participación mayoritaria de la comunidad
científica y de entusiastas de las emergentes tecnologías de telecomunicaciones. 

En la página 41, en la incompleta y reducida sección {}``Proyectos
de código abierto en desarrollo'', el informe dice textualmente:

\begin{quotation}
Parece muy claro que los usuarios finales no son la principal audiencia
de los desarrolladores.
\end{quotation}
A pesar de miles de grupos de usuarios de soporte, las listas de correos
para soporte y comentarios de usuarios%
\footnote{Una rápida visita a algunos de los sitios web de la mayoría de proyectos
de software libre servirá para convencer. Todas tienen enlaces a listas
de correo, sistemas de gestión de \emph{bugs}, documentación de programación
y de uso.%
} y el contacto directo de programadores con sus usuarios, hasta podríamos
otorgarles el beneficio de la duda. Pero esas funcionen las cumplen
los grupos de evaluación de usabilidad \cite{usa-kde,usa2} departamentos
de soporte de las empresas dedicadas al desarrollo, distribución o
soporte (IBM, HP, SuSE, RedHat, Mandrake, Ximian/Novell, Sun...).
¿O es que los programadores de Microsoft tienen un contacto más directo
con los usuarios finales?.

\subsection{\label{sub:El-gobierno-no}El gobierno no debe financiar a proyectos
software GPL}

El informe critica la intervención de la {}``mayoría de los gobiernos''
a través de {}``financiación sustancial'' (página 2, 75) a proyectos
de software con licencia GPL. No definen el significado cuantitativo
de {}``sustancial'' ni {}``mayoría''. Tampoco citan ni explican
las fuentes de donde han extraído esos datos. 

Dada la juventud relativa de la licencia GPL y de la formalización
del movimiento de {}``código abierto'', lo más probable es que la
inmensa mayoría de financiación en proyectos de software haya estado
dedicada al desarrollo de software propietario.

El informe cita en la misma página (75) el proyecto Beowulf desarrollado
en la NASA como ejemplo de software desarrollado bajo GPL y por lo
tanto {}``inservible'' para las empresas. Pero Beowulf \emph{no
es un programa, es un concepto}, factible de ser implementado sobre
prácticamente cualquier plataforma. La más popular es con librerías
de intercambios de mensaje MPI (\emph{Message Passing Interface})
o PVM (\emph{Parallel Virtual Machine}), ambas disponibles para plataformas
Windows y Unix, y bajo licencias BSD.

En la página 76 (nota al pie 234) también cita al proyecto de Hans
Reiser, el ReiserFS, un sistema de ficheros novedoso\cite{galli1}
desarrollado con financiación parcial de DARPA. Reiser libera todo
el código bajo licencia GPL, y esa es la razón principal de las criticas
del informe. La razón fundamental es que \emph{Microsoft no puede
apropiarse de dicho código para modificarlo y desarrollarlo como producto
propietario}%
\footnote{Aunque si son cuidadosos en cómo lo hacen, hasta podrían integrarlo
en sus sistemas operativos, como hizo TiVo, ATI o NVidia con sus módulos
propietarios.%
} \emph{sin respetar los estándares y formatos} originales%
\footnote{Ya existen precedentes con otros programas bajo licencia BSD, como
la pila TCP/IP y Kerberos, que son parte fundamental de los sistemas
de red y de autenticación de Microsoft.%
}. 

El deseo de Microsoft, expresado perfectamente en el informe, es que
los gobiernos sólo financien a proyectos de código abierto cuyos resultados
se liberen bajo licencia BSD o de {}``dominio público''. Así puedan
ser utilizados y modificados \emph{sin restricciones} por empresas
que no han participado en el trabajo.

La solicitud sería coherente si además solicitase que el desarrollo
de software propietario financiado por las administraciones siguiesen
la misma regla. Pero no lo hacen, es la expresión más elaborada de
\emph{lo tuyo es mío, lo mío es mío}.

Pero lo más interesante de las {}``evidencias'' y conclusiones del
informe son:

\begin{itemize}
\item \emph{Los gobiernos sólo deben tomar decisiones técnicas y económicas,
no políticas} (página 60). 
\item La sugerencia de que \emph{los gobiernos deberían tomar decisiones
basadas en principios de máxima ganancia}%
\footnote{¿Es un \emph{lapsus}? %
}\emph{, como las empresas} (página 67).
\item \emph{La legislación a favor del código abierto implica que hay algo
más que un simple debate técnico} (página 70).
\end{itemize}
En otras palabras, los gobiernos democráticos, con representantes
políticos elegidos que deben dirigir a estados independientes y soberanos,
no pueden ni deben debatir y tomar decisiones políticas. Sólo técnicas
y económicas, aunque sean temas que afectan a tecnologías e infraestructuras
básicas y transversales como la informática.

\section{Conclusiones}

Como se ha demostrado, el informe NERA no es un informe independiente.
Está realizado por encargo de Microsoft para servir de apoyo a sus
actividades de \emph{lobbying} para evitar leyes más favorables a
desarrollos de software no propietarios. Los autores del informe hace
años que trabajan estrechamente con Microsoft. Posiblemente gran parte
de la facturación de NERA provenga de Microsoft.

El informe, por supuesto, es favorable a Microsoft, pero está lleno
de falsedades y opiniones presentadas como si fuesen hechos estudiados
e irrefutables. Todo ello trufado con un lenguaje innecesariamente
complicado y técnico para confundir al lector y dar la impresión de
haber sido desarrollado y redactado por expertos documentados, a conciencia,
y siguiendo criterios científicos.

A lo largo de la exposición no han ahorrado esfuerzos en demonizar
al {}``código abierto'' en general y al {}``software libre'' en
particular mediante falsedades y críticas extremas a la licencia GPL
---creada por la \emph{Free Software Foundation} con el simple objetivo
de asegurar que las empresas que modifiquen, vendan o distribuyan
programas desarrollados por la comunidad devuelvan de alguna forma
el esfuerzo realizado por ella---.

El informe primero niega la evidencia que existan monopolios en la
industria informática. 

Luego justifica la existencia del mismo en aras de la innovación e
interés de los usuarios, que adjudican casi exclusivamente al software
propietario. Intenta justificar esta supuesta exclusividad en la innovación
mediante una más que controvertida extrapolación de la evolución del
número de patentes con un incremento de la innovación. A pesar de
los informes contrarios de especialistas y de los cambios sustanciales
en los criterios de evaluación de patentes de hace poco más de una
década.

A continuación critica que los políticos tomen decisiones políticas
y exige que se atengan a decisiones técnicas, como si dirigir un país
fuese igual que dirigir un departamento de informática de una pequeña
empresa sin ningún tipo de planificación a medio plazo.

Finalmente, y ante la previsión de que los estamentos políticos finalmente
decidan ayudar y financiar proyectos de código abierto, exige que
las licencias del software generado no sea GPL, sino preferiblemente
BSD o de {}``dominio público''. Por la sencilla razón de que de
esta forma las empresas desarrolladoras no sólo serán propietarias
de patentes que impedirán desarrollar aplicaciones en software libre,
sino que si se desarrolla algún proyecto novedoso en código abierto
lo podrán utilizar y comercializar sin la obligación de dar nada a
cambio al grupo que lo ha desarrollado. Ni a la sociedad que lo ha
financiado.

\begin{quote}
\bigskip{}
{\small La {}``producción de iguales'' }%
\footnote{{\small Se refiere al método de producción en el software de código
abierto.}%
} {\small nos presenta un fenómeno fascinante que podría permitirnos
canalizar importantes reservas de esfuerzos creativos infra utilizados.
Es de vital importancia que no lo ignoremos en los debates políticos,
o mejor aún, desplacemos sus beneficios a economías que sí lo valoren,
y aseguremos las condiciones institucionales que necesita para florecer.}{\small \par}

\begin{flushright}\textbf{\small Yochai Benkler} {\small \cite{penguin}}\end{flushright}{\small \par}
\end{quote}

\end{document}